\documentclass[twocolumn,showpacs,fleqn,nobibnotes]{revtex4-1}
\usepackage{float,subfig,graphicx,epsfig,epsf}
\usepackage{multirow}
\usepackage{url}
\usepackage{amsmath,units}
\usepackage[usenames,dvipsnames,svgnames,table]{xcolor}
\usepackage[colorlinks=true,linkcolor=red,urlcolor=blue,citecolor=blue]{hyperref}
\usepackage{comment}
\usepackage{textcomp}

\begin{document}

	\thispagestyle{empty}

\title{HEAVY MESONS PHOTOPRODUCTION IN PERIPHERAL AA COLLISIONS}
\pacs{12.38.Bx; 13.60.Hb}
\author{M. B. GAY DUCATI\footnote{beatriz.gay@ufrgs.br} AND S. MARTINS\footnote{sony.martins@ufrgs.br}}

\affiliation{High Energy Physics Phenomenology Group, GFPAE  IF-UFRGS \\
Caixa Postal 15051, CEP 91501-970, Porto Alegre, RS, Brazil}

\begin{abstract}
The exclusive photoproduction of the heavy vector mesons $\psi$'s and Y's is investigated in peripheral lead-lead collisions for the energies available at the LHC, $\sqrt{s}=2.76$ TeV and $\sqrt{s}=5.02$ TeV. In order to evaluate the robustness of the light-cone color dipole formalism, previously tested in the ultraperipheral regime, the rapidity distribution, and the nuclear modification factor ($R_{AA}$), are calculated for the three centrality classes: 30\%-50\%, 50\%-70\% and 70\%-90\%. The ultraperipheral to peripheral regime transition was carried out sophisticating the photon flux description and the photonuclear cross section, taking into account the effective interaction area. In our calculations, three scenarios were considered: (scenario 1) the direct application of the usual photon flux and of the photonuclear cross section with no relevant change in relation to the UPC’s; (scenario 2) the application of an effective photon flux keeping the photonuclear cross section unchanged; and (scenario 3) also considering an effective photonuclear cross section. The results obtained with the three different scenarios were compared with the ALICE measurements, showing a better agreement with the data (only $J/\psi$ at the moment) in the more complete approach (scenario 3), mainly in the more central regions (30\%-50\% and 50\%-70\%) where the incertainty is smaller.
\end{abstract}

 
  \maketitle
  
\section{Introduction}

One of the main contributions of HERA was the discovery of diffractive events, characterized by large rapidity gaps ($\eta\gtrsim 4$) with absence of hadronic activity, where one, or both hadrons emerge intact in the final state, represent a relevant fraction of deep inelastic scattering \cite{arxiv0511047}. Soft diffraction events contribute with ($\sim$ 20\%) of the total inelastic proton-proton cross section and therefore must be taken into account in order to keep the background of many processes in the LHC \cite{EPJC54199, AIPCP1038-95} under control. The hard diffractive processes, responsible for the production of the states with high mass, or high $p_T$ (ex. $Q\bar{Q}$, jets, W, Z), can be calculated from perturbative QCD. The presence of a hard scale allows us to obtain more information about the Diffractive (or Generalized) Particle Distribution Functions, which describe not only the particle density, but also the correlation between them \cite{PR38841}. In addition, the perturbative treatment also opens a way to test new ideas on the mechanism of exchange of two or more gluons in a singlet color state (Pomeron IP).

An interesting type of diffractive process is the exclusive photoproduction of the vector mesons, in which the collision of two hadrons produces a vector meson, keeping intact the initial hadrons. This mechanism is dominant in the ultraperipheral regime, and the cross section is factorized in two terms: a quasi-real photon flux, created from one of the incoming hadrons, and the photoproduction cross section, that characterizes the interaction of the photons with the target-hadron. The exclusive photoproduction has been investigated in several works \cite{PRD94-094023,PRD88-017504,JPG42-105001,PRC87-032201,PRC93-055206,AIP1654,TMP182-141}. In one of our last contributions to the subject \cite{PRD94-094023}, we calculated the rapidity distribution for the production of Y's states in Pb-Pb collisions at energies $\sqrt{s} = 2.76$ TeV and $\sqrt{s} = 5.02$ TeV. In that case, the light cone colour dipole formalism was used \cite{nik}, including the partonic saturation and the nuclear shadowing effects \cite{NPB268-427,PLB226-167,NPB493-305,NPB511-355}. This time, we are interested in testing the robusteness of the dipole colour formalism in the peripheral regime. In this region, the ALICE and STAR collaborations measured an excess in the $J/\psi$ production in small $p_T$ \cite{PRL116-222301,JPCS779-012039}, which could be the product of exclusive photoproduction. There are very few studies dealing with this production mechanism in the peripheral collisions regime \cite{PRD96-056014,PRC93-044912,PRC97-044910}. In \cite{PRC93-044912}, that also motivates the present study, this issue is treated by modifying one of the components of the cross section, the photon flux. However, no change is made in the photoproduction cross section in relation to the ultraperipheral case. In our first paper related to the subject \cite{PRD96-056014}, it was calculated the rapidity distribution of the $J/\psi$ with an effective photon flux constructed in terms of the usual photon flux. An effective b-dependent interaction area was considered instead of a constant value $\pi R_A^2$ adopted in \cite{PRC93-044912}. Here, these previous calculations are refined applying a relevant modification also in the photoproduction cross section, following the geometrical formalism adopted in the construction of the effective photon flux. In relation to the last work, it was included a new centrality class (30\% -50\%) as well as another dipole model (IIM model). It was also enlarged the number of states under analysis, calculating the rapidity distribution for the mesons $\psi(2S)$ and $Y(1S,2S,3S)$ to produce a more comprehensive analysis. The main goal of this work is to evaluate the behavior of the nuclear modification factor in three different approaches: (scenario 1) direct application of the usual photon flux and of the photonuclear cross section without any modification in relation to UPCs,(scenario 2) altering only in the photon flux, and (scenario 3) modifying the photon flux and in the photonuclear cross section. 
	
\section{Experimental Approach}\label{exp}

\begin{figure}[H]
\centering
\scalebox{0.6}{
\includegraphics{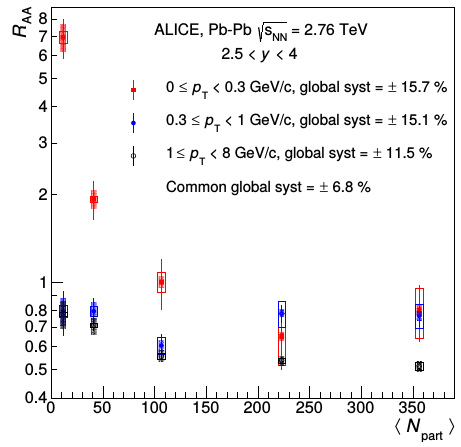}}
\caption{$R_{AA}$ for $J/\psi$ production as a function of the average number of participating nucleons $\langle N_{\text{part}}\rangle$. Figure extracted from \cite{PRL116-222301}.\label{figraa}}
\end{figure}

The ALICE and STAR collaborations measured the peripheral hadroproduction of the $J/\psi$ in AA collisions, revealing an excess in the production of this meson in the small transverse momentum ($p_T<0.3$ GeV/c) at forward ($2.5<y<4.0$) and mid-rapidity ($|y|<1$), respectively. In the ALICE paper \cite{PRL116-222301}, the average radidity distribution and the nuclear modification factor $R_{AA}$ for Pb-Pb collisions were explored in the 30\%-50\%, 50\%-70\% and 70\%-90\% centrality classes at $\sqrt{s}=2.76$ TeV, Fig. \ref{figraa}. In addition, in the STAR paper \cite{JPCS779-012039}, the $J/\psi$ invariant yield and $J/\psi$ $R_{AA}$ were measured as a function of $p_T$ for 20\%-40\%, 40\%-60\% and 60\%-80\% centrality classes at $\sqrt{s}=200$ GeV (Au-Au) and $\sqrt{s}=1.93$ GeV (U-U). In this work, the main goal is to test the robustness of the color dipole formalism in the same energy limit used in our previous works ($\sqrt{s}=2.76$ TeV and $\sqrt{s}=5.02$ TeV) to describe the rapidity distribution and $R_{AA}$. Thus, the results produced in this work will be compared with ALICE data. In a further study focused on $p_T$ dependence, data from both experiments should be considered.

To estimate the $R_{AA}$ values, it was adopted the expression developed in \cite{PLB734-314}, 
\begin{eqnarray}
\scalebox{0.75}{$R_{AA}^{hJ/\psi}=\dfrac{N_{AA}^{J/\psi}}{BR_{J/\psi\rightarrow l^+l^-}\cdot N_{events}\cdot\left(A\times\varepsilon\right)_{AA}^{J/\psi}\cdot\left\langle T_{AA}\right\rangle \cdot\sigma_{pp}^{hJ/\psi}}$}\label{eq:raa},
\end{eqnarray}
where, $N_{AA}^{J/\psi}$ represents the measured number of the $J/\psi$ ($N_{AA}^{J/\psi}$). This number was then corrected for the acceptance times efficiency $\left(\mathcal{A}\times\varepsilon\right)_{AA}^{J/\psi}\sim 11.31\%$ taking into account that photoproduced $J/\psi$ are expected to be transversally polarized, for the branching ratio $\text{BR}_{J/\psi\rightarrow \mu^+\mu^-}=5.96$\%, normalized to the number of equivalent MB events $N_{events}\simeq 10.6\times 10^7$ (calculated from \cite{PLB734-314}), for the average nuclear overlap function $\left(\langle T_{AA}\rangle\right)$, which values depend on the centrality class, and can be calculated from the Table I of \cite{PRC88-044909}. In the case of the classes (30\%-50\%), (50\%-70\%) and (70\%-90\%), we obtained the values $3.84\text{ mb}^{-1}$, $0.954\text{ mb}^{-1}$ and $0.17\text{ mb}^{-1}$, respectively. At last, the result is normalized by proton-proton inclusive $J/\psi$ production cross section, ($\sigma_{pp}^{hJ/\psi}$), which is calculated using the parametrization suggested in \cite{PRL116-222301}, where
\begin{eqnarray}
\dfrac{d^{2}\sigma_{pp}^{hJ/\psi}}{dp_{T}dy}=\dfrac{c\cdot\sigma_{J/\psi}\cdot p_{T}}{1.5\cdot\left\langle p_{T}\right\rangle ^{2}}\left(1+a^{2}\left(\dfrac{p_{T}}{\left\langle p_{T}\right\rangle }\right)^{2}\right)^{-n}\label{eq:fit}
\end{eqnarray}
with 
\begin{eqnarray}
a=\frac{\Gamma\left(3/2\right)\Gamma\left(n-3/2\right)}{\Gamma\left(n-1\right)}\qquad c=2a^{2}\left(n-1\right).\nonumber
\end{eqnarray}
The values of the free parameters $\sigma_{J/\psi}$, $\left\langle p_{T}\right\rangle$ and $n$ are obtained from the fit of the equation (\ref{eq:fit}) with the ALICE data at large $p_T$ \cite{PLB718-295}. This procedure results in $\sigma_{J/\psi} = 3.31$, $\left\langle p_{T}\right\rangle = 2.369$ and $n = 4.76$. Applying these values in (\ref{eq:fit}) and taking the integral in the kinematical regions of interest, $p_{T}<0.3\text{ GeV}$ and $2.0<y<4.5$, it is obtained $\sigma_{pp}^{hJ/\psi}=0.0514\text{ }\mu\text{b}$. 

In the ALICE measurement, both production mechanisms (hadro and photo) are considered, being unable the measurement of each component. Therefore, the number of the $J/\psi$ events should be separated in two terms
\begin{eqnarray}
N_{AA}^{J/\psi}=\underset{\text{hadro}}{\underbrace{N_{AA}^{hJ/\psi}}}+\underset{\text{photo}}{\underbrace{N_{AA}^{\gamma J/\psi}}}.
\end{eqnarray}
For the hadro component, the $J/\psi$ hadroproduction was not directly calculated, but estimated from an analyses performed on the ALICE data, as shown in Figure \ref{figraa}. Taking the centrality 70\%-90\% centrality, for example, the $R_{AA}$ measured value in the region $p_T<0.3$ GeV is $7$, while in the range $1.0<p_T<8.0$, it is approximately, $0.8$. The $R_{AA}$ for $p_T>1$ GeV gives a good approximation for the hadronic production of the $J/\psi$; and assuming that this contribution stays the same in the $p_T<0.3$ region, it is possible to calculate the proportion between the hadro and photo contributions. Although the $R_{AA}^{hJ/\psi}$ data were obtained from a parametrization like Woods-Saxon, with $p_T$ dependence \cite{PRL116-222301}, the rough parametrizations $R_{AA}^{hJ/\psi}\left(p_T<0.3\text{ GeV/c}\right)=R_{AA}^{hJ/\psi}\left(1<p_T<8\text{ GeV/c}\right)$ and $R_{AA}^{hJ/\psi}\left(p_T<0.3\text{ GeV/c}\right)=1$, as discussed in \cite{PRL116-222301}, reafirm the presence of the excess of $J/\psi$ compatible with the results reported in the experimental paper. Thus, the hypothesis considered here produces $0.8/7\sim 0.11$ for 70\%-90\% centrality class. Repeating the same procedure for the other centrality classes, 30\%-50\%, and 50\%-70\%, it gives 
\begin{eqnarray*}
\left(R_{AA}^{hJ/\psi}\right)^{30-50} & \sim & 0.56\left(R_{AA}^{\gamma J/\psi}\right)^{30-50}\\
\left(R_{AA}^{hJ/\psi}\right)^{50-70} & \sim & 0.36\left(R_{AA}^{\gamma J/\psi}\right)^{50-70}\\
\left(R_{AA}^{hJ/\psi}\right)^{70-90} & \sim & 0.11\left(R_{AA}^{\gamma J/\psi}\right)^{70-90}
\end{eqnarray*}
such that, 
\begin{eqnarray*}
N_{AA}^{J/\psi}=\begin{cases}
2.27N_{AA}^{\gamma J/\psi} & \qquad\text{for}\quad30\%-50\%\\
1.56N_{AA}^{\gamma J/\psi} & \qquad\text{for}\quad50\%-70\%\\
1.12N_{AA}^{\gamma J/\psi} & \qquad\text{for}\quad70\%-90\%
\end{cases}
\end{eqnarray*}

Comparing the central values of the excess of $J/\psi$ with the average rapidity distribution, both shown in the Table I of the \cite{PRL116-222301}, we can infer the approximate relation \scalebox{0.7}{$N_{AA}^{\gamma J/\psi}\sim0.86\cdot10^{6}\dfrac{d\sigma_{J/\psi}^{\gamma}}{dy}$}, which is valid for the three centrality classes of interest. Then, the calculation of the $N^{J/\psi}_{AA}$ sets just in the calculation of the average rapidity distribution, which is estimated from the integration in the range $2.5<y<4.0$, as 
\begin{eqnarray}
\scalebox{0.7}{$\left.\dfrac{d\sigma_{J/\psi}^{\gamma}}{dy}\right|_{2.5<y<4.0} = \dfrac{\int\dfrac{d\sigma_{J/\psi}^{\gamma}}{dy}dy}{\Delta y}$}\label{dsidymed} 
\end{eqnarray}
The rapidity distribution is calculated in this work employing the photon flux and  the colour dipole formalism, which are detailed in the next section. 

\section{Theoretical Framework}\label{ipdpf}

In the ultrarelativistic limit, the exclusive nuclear photoproduction cross section of the vector meson $V$ can be written as the product of a quasi-real photon flux, which is produced from one of the nucleus, and the photonuclear cross section that corresponds to the photon-nuclei interaction $\gamma A\rightarrow V+A$ \cite{ARNPS55-271}. Considering that the photon flux carries the dependence with the impact parameter of the collision $b$, the differential cross section in the rapidity $y$ and in the impact parameter $b$ can be defined by \cite{PRC93-044912}
\begin{eqnarray}
\scalebox{1.0}{$\dfrac{d^3\sigma_{AA\rightarrow AAV}}{d^2bdy} = \omega N(\omega,b)\sigma_{\gamma A\rightarrow VA}+\left(y\rightarrow -y\right)$}\label{pri}.
\end{eqnarray}
where $\omega=\frac{1}{2}M_V\text{exp}(y)$ is the photon energy and $M_V$ is the meson mass. 

The photon flux espectra, $N(\omega,b)$, is directly connected with the eletromagnetic distribution of the emitting nucleus, and is described by the nuclear form factor $F(k^2)$, which is the Fourier transform of the nuclear density profile. To ensure the dependence of the photon flux on the form factor, the generic formula presented in \cite{PPNP39-503} is used, 
\begin{eqnarray}
\scalebox{1.0}{$N\left(\omega,b\right)=\dfrac{Z^{2}\alpha_{QED}}{\pi^{2}\omega}\left|\int_{0}^{\infty}dk_{\perp}k_{\perp}^{2}\dfrac{F\left(k^2\right)}{k^2}J_{1}\left(bk_{\perp}\right)\right|^{2}$}\label{eq:xx},
\end{eqnarray}
where $Z$ is the nuclear charge, $\gamma=\sqrt{s_{NN}}/(2m_{\textrm{proton}})$ is the Lorentz factor, $k_{\perp}$ is the transverse momentum of the photon and $k^{2}=\left(\omega/\gamma\right)^{2}+k_{\perp}^{2}$. For a heavy nuclei as gold or lead, the Fermi distribution with 2 parameters (sometimes called Woods-Saxon) is more suitable. However, to obtain the analytic result of the form factor from this distribution is unlikely, requiring the adoption of the approximation shown in \cite{PRC14-1977,PRC60-014903}, where the Woods-Saxon distribution is rewritten as a hard sphere, with radius $R_A$, convoluted with an Yukawa potential with range $a=7$ fm. The Fourier transform of this convolution is the product of the individual transformation as
\begin{eqnarray}
\scalebox{1.0}{$F(k)$}&\scalebox{1.0}{$=$}&\scalebox{1.0}{$\dfrac{4\pi\rho_{0}}{Ak^{3}}\left[\textrm{sin }\left(kR_{A}\right)-kR_{A}\textrm{cos }\left(kR_{A}\right)\right]$}\nonumber
\\
&\scalebox{1.0}{$\times$}&\scalebox{1.0}{$\left[\dfrac{1}{1+a^{2}k^{2}}\right]$},\label{eq:wsy}
\end{eqnarray}
where $A$ is the mass number of the nuclei and $\rho_{0}=0.1385\textrm{ fm}^{-3}$ for lead.

In Eq. (\ref{pri}), the $\sigma_{\gamma A\rightarrow VA}$ is the coherent photonuclear cross section, which characterizes the photon-nuclei. In accordance with the paper \cite{PRC85-044904}, the cross section for the photoproduction of a vector meson V on H (H $\equiv$ p, A) can be factorised in two components: the forward scattering amplitude ($d\sigma/dt|_{t=0}$), which carries the dynamical information of the process and the form factor, $F(t)$, which is, in general, dependent on the spatial characteristics of the target. This factorization has being commonly used in the literature as can be seen in the works \cite{EPJC40-519, PRC93-044912, PRC85-044904}, where the forward scattering amplitude was characterized, respectively, by the vector meson dominance,  perturbative QCD and color dipole formalisms. Thus, the coherent photonuclear cross section is defined as
\begin{eqnarray}
\scalebox{0.8}{$\sigma_{(\gamma A\rightarrow VA)}=\frac{|\textrm{Im }A(x,t=0)|^2}{16\pi}\left(1+\beta^2\right)R^2_g\int_{t_{min}}^{\infty}|F(t)|^2dt$},\label{seg}
\end{eqnarray}
The parameter  $\beta=\mathrm{Re}A/\mathrm{Im}A$ restores the real contribution of the scattering amplitude and is usually defined as \cite{PRD74-074016}
\begin{eqnarray}
\scalebox{0.8}{$\beta = \textrm{tan }\left(\frac{\pi\lambda{eff}}{2}\right)\textrm{ , where }\lambda_{eff}=\dfrac{\partial\textrm{ln }\left[\textrm{Im }A(x,t=0)\right]}{\partial\textrm{ln }s}$}\nonumber.
\end{eqnarray}
Other important parameter, $R_g^2(\lambda_{eff})$, is necessary for heavy mesons as $J/\psi$, and corresponds to the ratio of off-forward to forward gluon distribution (skewedness effect), being defined by \cite{PRD60-014015}
\begin{eqnarray}
R_g^2(\lambda_{eff})=\dfrac{2^{2\lambda_{eff}+3}}{\sqrt{\pi}}\dfrac{\Gamma\left(\lambda_{eff}+\frac{5}{2}\right)}{\Gamma\left(\lambda_{eff}+4\right)}\nonumber
\end{eqnarray}
The function $F(t)$ is the same nuclear form factor shown in (\ref{eq:wsy}), which is integrated from $t_{min}=\left(M_V^2/2\omega\gamma\right)^2$.

At last, the amplitude $\left|\text{Im }A(x,t=0)\right|$ represents the imaginary part of the interaction amplitude for the $\gamma A\rightarrow V+A$ process. Based in good results obtained in last works \cite{PRD94-094023,PRD88-017504,JPG42-105001}, we described the amplitude $\textrm{Im }A(x,t=0)$ in the colour dipole formalism, where the photon-nuclei scattering can be seen as a sequence of the following subprocesses: (i) the photon fluctuates into a quark-antiquark pair (the dipole), (ii) the dipole-target interaction and (iii) the recombination of the $q\bar{q}$ into a vector meson. In the quantum mechanical picture of diffraction developed by Good and Walker \cite{PR120-1857}, the amplitude of this sequence of steps is written in the dipole formalism as
\begin{eqnarray}
\scalebox{0.9}{$\textrm{Im }A(x,t=0)=\int d^2r\int\frac{dz}{4\pi}\left(\psi^*_V\psi_{\gamma}\right)_T\sigma_{\textrm{dip}}^{\textrm{nucleus}}\left(x,r\right)$},\label{ima}
\end{eqnarray}
where the variables $z$ and $r$ are the longitudinal momentum fraction carried by the quark and the transverse color dipole size, respectively. The Eq. (\ref{ima}) is safely applicable in the low-x limit, in which the transverse size of the pair $q\bar{q}$ is frozen during the interaction with the target ensuring the applicability of the dipole formalism. It is not formally defined where the low-x limit starts and, in this work, we extended the formalism up to $y=4$ ($x \sim 0.06$ for $J/\psi$), which is a limit value commonly used in the UPC regime. The saturation model is more suitable in the region below $x=0.01$ and for large $x$ limit, there is still need of a complete analytical treatment. However, the photon flux at $y = 4$ corresponds to photons with energy $\sim$ 84 GeV (for $J/\psi$), which are strongly suppressed in relation to photons with energy $\lesssim$ 0.2 GeV. Meaning that possible corrections to the dipole formalism for higher values of $x$ would be suppressed by the photon flux.

The transverse overlap of the photon-meson wave function, which is dominant in relation to longitudinal component in $Q\sim0$, can be written as \cite{PRD74-074016}
\begin{eqnarray}
\scalebox{0.9}{$\left(\psi^*_V\psi_{\gamma}\right)_T$} &\scalebox{0.9}{$=$}& \scalebox{0.9}{$\hat{e}_fe\frac{N_c}{\pi z(1-z)}\left\{m^2_fK_0(\epsilon r)\phi_T(r,z)\right.$}\nonumber
\\
&\scalebox{0.9}{$-$}&\scalebox{0.9}{$\left.\left[z^2+(1-z)^2\right]\epsilon K_1(\epsilon r)\partial_r\phi_T(r,z)\right\}$}
\end{eqnarray}
where $\hat{e}_f=1/3$ for $J/\psi$, $e=\sqrt{4\pi\alpha_{em}}$, $\epsilon^2=z(1-z)Q^2+m^2_f$ and $N_c=3$. The phenomenological function $\phi_T(r,z)$ represents the scalar part of the meson wave-function and, here, it was used the Boosted-Gaussian model \cite{ZPC75-71}, since it can be applied in a systematic way for the excited states, resulting in
\begin{eqnarray*}
\phi_{1S}(r,z) &=& G_{1S}(r,z)
\\
\phi_{2S}(r,z) &=& G_{2S}(r,z)\left[1+\alpha_{2S,1}g_{2S}(r,z)\right]
\\
\phi_{3S}(r,z) &=& G_{3S}(r,z)\left\{1+\alpha_{3S,1}g_{3S}(r,z)+\alpha_{3S,2}\right.
\\
&\times&\left.\left[g_{3S}^2(r,z)+4\left(1-\frac{4z(1-z)r^2}{R_{3S}^2}\right)\right]\right\}
\end{eqnarray*}
where
\begin{eqnarray}
\scalebox{0.8}{$G_{nS}(r,z)=\mathcal{N}_{nS}z(1-z)\text{exp}\left(-\frac{m_{c/b}^2\mathcal{R}^2_{nS}}{8z(1-z)}-\frac{2z(1-z)r^2}{R^2_{nS}}+\frac{m_{c/b}^2\mathcal{R}^2_{nS}}{2}\right)$}\nonumber
\end{eqnarray}
and
\begin{eqnarray}
\scalebox{0.8}{$g_{nS}(r,z)=2-m_{c/b}^2\mathcal{R}^2_{nS}+\frac{m_{c/b}^2\mathcal{R}^2_{nS}}{4z(1-z)}-\frac{4z(1-z)r^2}{R^2_{nS}}$}\nonumber.
\end{eqnarray}
The free parameters $\mathcal{R}^2_{nS}$, $\mathcal{N}_{nS}$ and $\alpha_{nS}$ are determined from normalization, the orthogonality conditions, and a fit to the experimental leptonic decay width (more details are found in \cite{PRD90-054003,Sanda2} where the parameters are calculated).

The next term in the equation (\ref{ima}) is the cross section $\sigma_{\textrm{dip}}^{\textrm{nucleus}}\left(x,r\right)$, calculated via Glauber model \cite{EPJC26-35}, 
\begin{eqnarray}
\scalebox{0.9}{$\sigma_{\textrm{dip}}^{\textrm{nucleus}}\left(x,r\right)$}&\scalebox{0.9}{$=$}&\scalebox{0.9}{$2\int d^2b$}\nonumber
\\
&\scalebox{0.9}{$\times$}&\scalebox{0.9}{$\left\{1-\textrm{exp}\left[-\frac{1}{2}T_A(b)\sigma_{\textrm{dip}}^{\textrm{proton}}\left(x,r\right)\right]\right\}$}\label{glauber}
\end{eqnarray}
where the nuclear profile function, $T_A(b)$, is obtained from a 2-parameter Fermi distribution for the nuclear density \cite{Fermi} and the dipole cross section, $\sigma_{\textrm{dip}}^{\textrm{proton}}(x,r)$, is modeled from the GBW \cite{PRD59-014017} and IIM \cite{PLB59-199} models, since both models presented good results in the ultraperipheral regime \cite{PRD94-094023,PRD88-017504,JPG42-105001}.

Considering the kinematical range $2.5<y<4.0$ and $p_T<0.3$ GeV, we combined the usual photon flux (eq. (\ref{eq:xx})) with the photoproduction cross section (eq. (\ref{seg})) to calculate the average rapidity distribution as described in the section \ref{exp}. To take into account the centrality range, the relation $c=b^2/{4R_A}$ was used, where $c$ corresponds to centrality. Integrating in the impact parameter for the centrality classes (30\%-50\%), (50\%-70\%) and (70\%-90\%), the results shown in the Table \ref{tab:dsigdypoor} are obtained, which also presents the ALICE data. It is observed an excellent agreement with the data in the more peripheral region. In contrast, as going to more central regions, our estimates overstimate the ALICE data and, therefore, some correction with $b$ dependence is required, as will be develped in the next section.

\begin{table}[H]
\centering
\scalebox{0.7}{
\begin{tabular}{|c|c|c|c|}
\multicolumn{4}{c}{Average Rapidity Distribution - scenario 1}\tabularnewline
\hline
$d\sigma/dy$ [$\mu$b] & 30\%-50\% & 50\%-70\% & 70\%-90\%\tabularnewline
\hline 
GBW & 200 & 100 & 60\tabularnewline
\hline 
IIM & 170 & 84 & 51\tabularnewline
\hline 
ALICE data & $73\pm44^{+26}_{-27}\pm10$ & $58\pm16^{+8}_{-10}\pm8$ & $59\pm11^{+7}_{-10}\pm8$\tabularnewline
\hline 
\end{tabular}}
\caption{Comparison of results for the $d\sigma/dy$ using GBW and IIM models with the ALICE data for $J/\psi$ \cite{PRL116-222301}. \label{tab:dsigdypoor}}
\end{table}

\section{The Effective Photon Flux}

\begin{figure}[H]
  \centering
  \scalebox{0.7}{
  \includegraphics{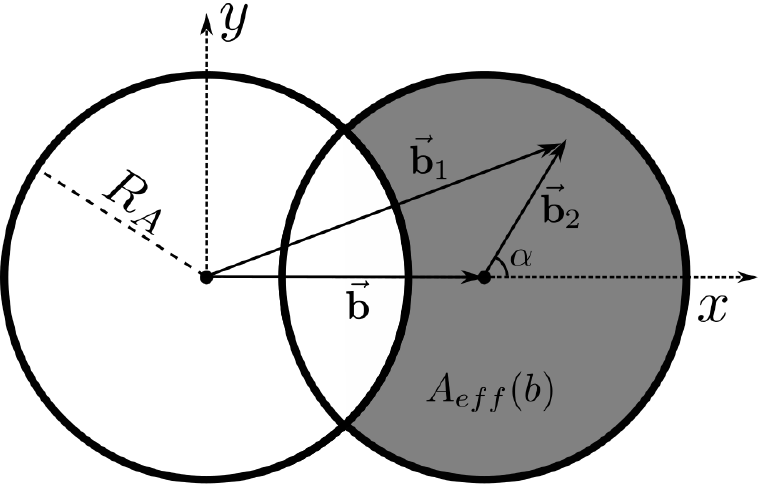}}
  \caption{Schematic drawing used in the construction of the effective photon flux. \label{fig:N2central}}
\end{figure}

To improve the calculations, the photon flux is modified following a similar procedure to the one carried out in \cite{PRC93-044912}, in which an effective photon flux is built as a function of the usual photon flux with two restrictions: (1) only photons that reach the geometrical region of the nuclei-target are considered, and (2) the photons that reach the overlap region are desconsidered. Consequentely, the vector $\vec{b}_1$ that starts in the center of the flux emitter nuclei will map only the allowed region of the target-nuclei (shaded region of the Figure \ref{fig:N2central}) . In contrast with \cite{PRC93-044912}, we do not divide by a fixed region $\pi R_A^2$. Being interested in collisions with centrality which extends from 30\% to 90\%, it is required to divide by the mapped areas, $A_{eff}(b)$, enforcing $b$ dependence. Thus, obtaining 
\begin{eqnarray}
\scalebox{0.7}{$N^{eff}\left(\omega,b\right)=\frac{1}{A_{eff}(b)}\int d^{2}b_{1}N\left(\omega,b_{1}\right)\theta(R_{A}-b_{2})\theta(b_{1}-R_{A})$}\label{eq:fluxoefetivo}
\end{eqnarray}
where 
\begin{eqnarray}
\scalebox{0.9}{$A_{eff}(b)=R_A^2\left[\pi-2\textrm{cos}^{-1}\left(\frac{b}{2R_A}\right)\right]+\frac{b}{2}\sqrt{4R_A^2-b^2}$}\nonumber.
\end{eqnarray}

The first condition to ensure that the effective photon flux is referred only to photons that reach the geometrical region of the nuclei-target, is given by the function $\theta\left(R_{A}-b_{2}\right)$. The equation (\ref{eq:fluxoefetivo}) can be correspondingly rewritten as 
\begin{eqnarray}
\scalebox{0.9}{$N^{eff}\left(\omega,b\right)=\frac{1}{A_{eff}(b)}\int b_2db_2d\alpha N\left(\omega,b_{1}\right)\theta(b_{1}-R_{A})$}\label{Neff1}\textrm{.}
\end{eqnarray}

Here the flux is expressed in terms of the new variable $\left(b_2, \alpha\right)$ related to $b_1$ through the equation $b_1^2=b^2+b_2^2+2bb_2\text{cos}(\alpha)$. The $\theta(b_{1}-R_{A})$ function corresponds to the second condition and, therefore, discards the contribution of the photons that reach the overlap region where the nuclear effects are present. With this last condition, the equation (\ref{Neff1}) can be separated in two components,
\begin{eqnarray}
\scalebox{0.9}{$N^{eff}\left(\omega,b\right)=\frac{1}{A_{eff}(b)}\left[N^{eff}_{full}\left(\omega,b\right)-N^{eff}_{overlap}\left(\omega,b\right)\right]$}\label{Neff}.
\end{eqnarray}

The first term, $N^{eff}_{full}\left(\omega,b\right)$, maps all the nuclear region, including the overlap region, 
\begin{eqnarray*}
\scalebox{0.9}{$N^{eff}_{full}\left(\omega,b\right)=\int b_2db_2d\alpha N\left(\omega,b_{1}\right)$};
\end{eqnarray*}
the second term, $N^{eff}_{overlap}$, maps only the overlap region and its contribution is defined in the cartesian coordinate system by 
\begin{eqnarray}
\scalebox{0.9}{$N^{eff}_{overlap}\left(\omega,b\right) = 2\int_{0}^{b_{ymax}}db_y\int_{b_{xmin}}^{b_{xmax}}db_xN\left(\omega,b_1\right)$}\label{nover},
\end{eqnarray}
where $b_1^2=b_x^2+b_y^2$, and the integration limits are \scalebox{0.7}{$b_{xmin}=-\sqrt{R_A^2-b_y^2}+b$}, \scalebox{0.7}{$b_{ymax} = \sqrt{R^2_A-\left(\frac{b}{2}\right)^2}$} and \scalebox{0.7}{$b_{xmax}=\sqrt{R_A^2-b_y^2}$.} For purposes of numerical calculation, it is more useful to disconnect the dependence of the $b_{xmin}$ and $b_{xmax}$ with $b_y$. This is achieved with the change of variables $b_y = g_1\left(b\right)b_y'$ and $b_x = g_2\left(b,b_y\right)b_x'+b/2$, where the Jacobian functions $g_{1,2}$ are respectively given by \scalebox{0.7}{$g_1\left(b\right) = \sqrt{R_A^2-\left(b/2\right)^2}$} and \scalebox{0.7}{$g_2\left(b,b_y\right) = \left(\sqrt{R_A^2-b_y^2}-b/2\right)$}. In terms of the new variables, the Eq. (\ref{nover}) can be rewritten as
\begin{eqnarray}
\scalebox{0.8}{$N^{eff}_{overlap}\left(\omega,b\right) = 2\int_0^1\int_{-1}^1 db_y'db_x'g_1\left(b\right)g_2\left(b,b_y\right)N\left(\omega,b_{1}\right)$}.
\end{eqnarray}

\begin{figure}[H]
  \centering
  \scalebox{0.4}{
  \includegraphics{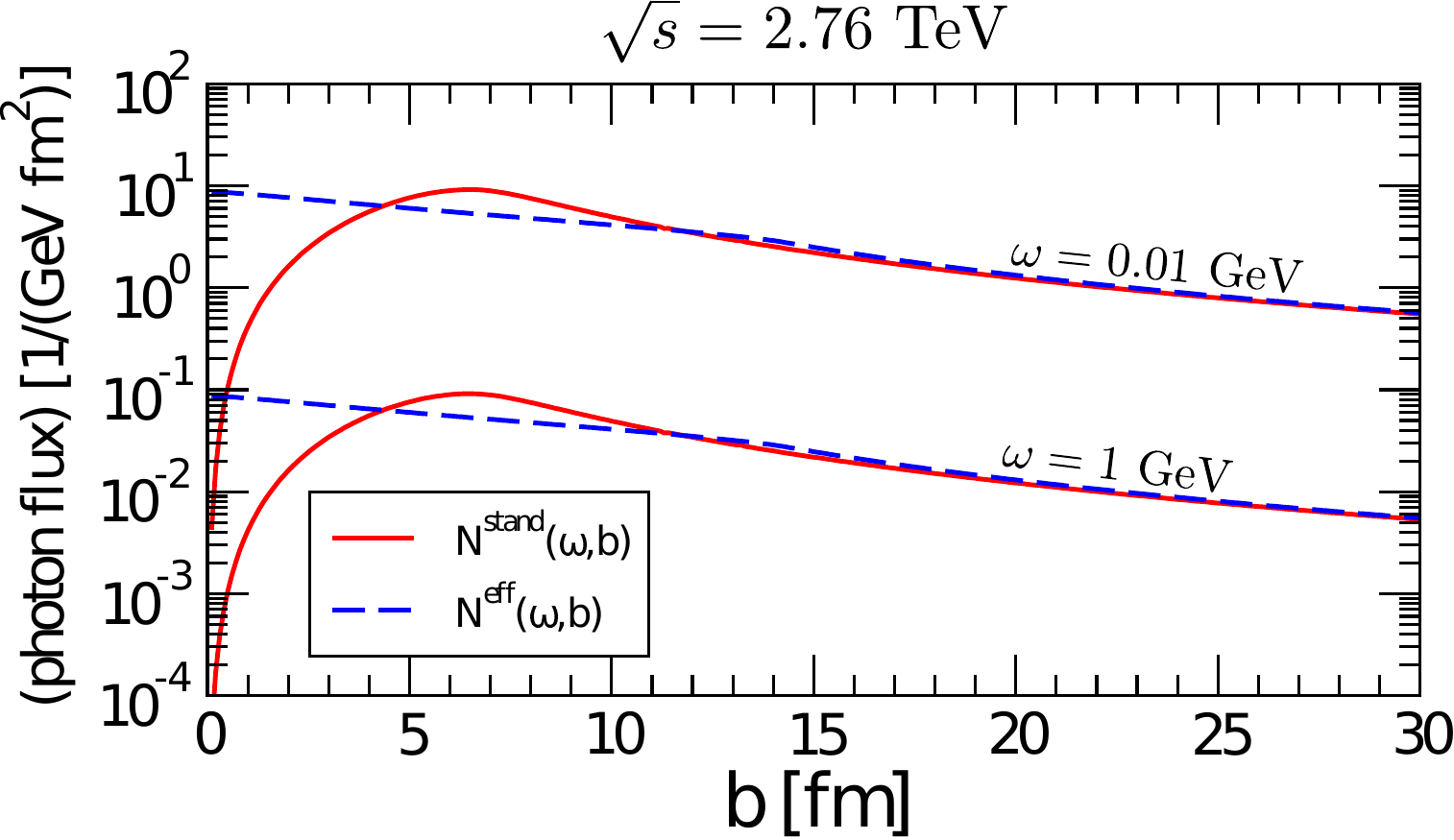}}
  \caption{Comparison between the usual photon flux (solid line) and the effective photon flux (dashed line) for the photon energy values $\omega=0.01$ GeV and $\omega=1$ GeV. \label{fig:fig3}}
\end{figure}

The Figure \ref{fig:fig3}, presents the comparison of the usual photon flux (Eq. (\ref{eq:xx})) with the effective photon flux (Eq. (\ref{Neff})) for the energies $\omega=0.01$ GeV and $\omega=1$ GeV, since for the centrality class 30\%-90\%, the photon flux is formed mainly for photons with energy $\omega<200$ MeV. For $b\lesssim 4$ fm (centrality $\lesssim$ 8\%), the usual photon flux considerably diverges from the effective photon flux, tending to 0 as $b\rightarrow 0$. Otherwise, in the range $4\text{ fm}\lesssim b\lesssim 11\text{ fm}$ ($8\% \lesssim\text{centrality}\lesssim 60\%$), the usual photon flux is higher than the effective photon flux, mainly in the limit $b\sim R_A\sim 7$ fm. At last, in the region $b>11$ fm, the two models become similar for the energy $\omega = 0.01$ GeV and $\omega = 1$ GeV, approaching each other as we reach the ultraperipheral regime ($b>2R_A$). 

\begin{figure}[h]
\centering
\scalebox{0.55}{
\includegraphics{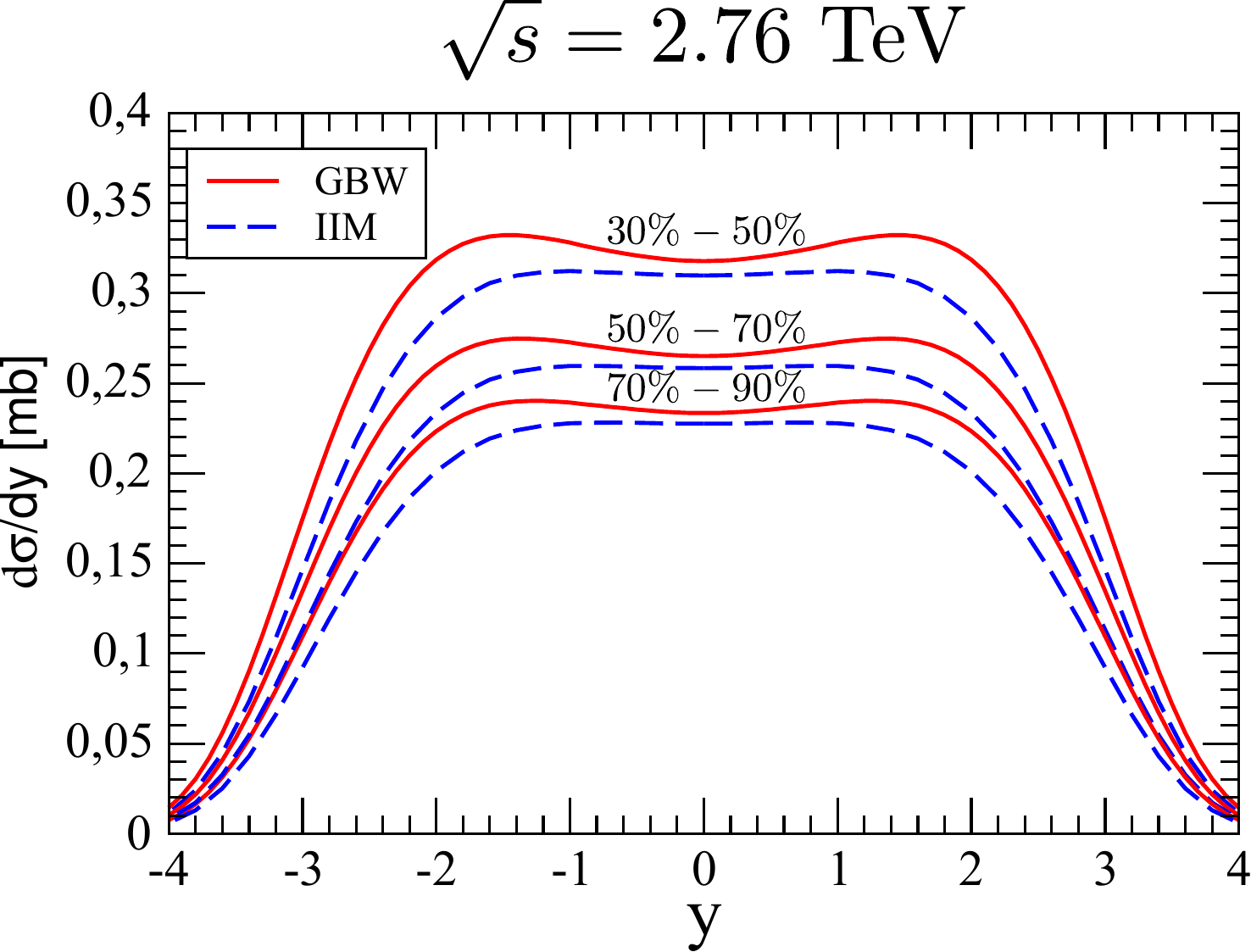}}
\caption{Rapidity distribution for $J/\psi$ nuclear photoproduction at $\sqrt{s}=2.76$ TeV for different centrality classes using the GBW and IIM dipole models.}
\label{psi1s}
\end{figure}

\begin{figure}[h]
\centering
\scalebox{0.55}{
\includegraphics{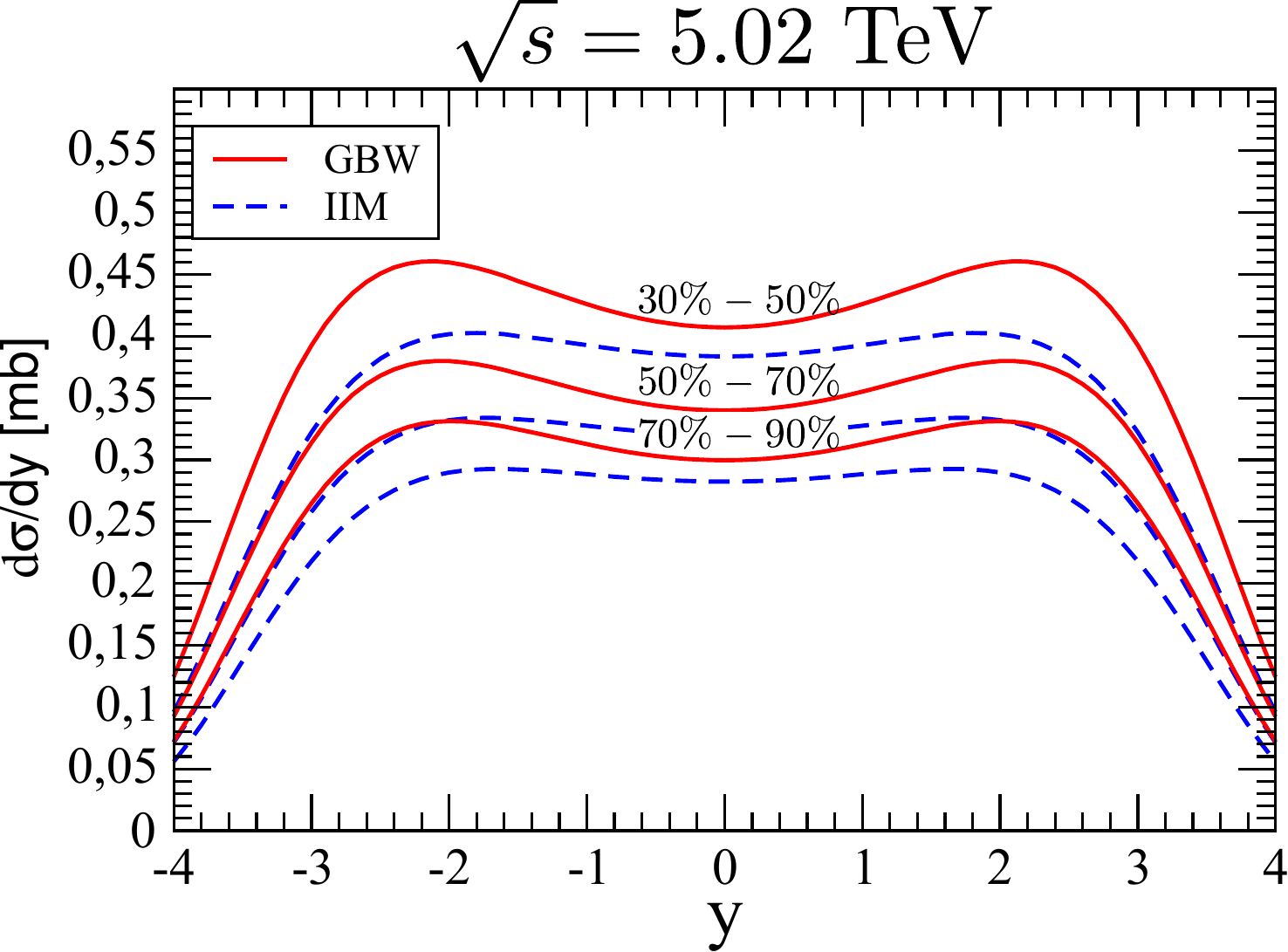}}
\caption{Rapidity distribution for $J/\psi$ nuclear photoproduction at $\sqrt{s}=5.02$ TeV for different centrality classes using the GBW and IIM dipole models.}
\label{psi1s5020}
\end{figure}

Using the effective photon flux (Eq. (\ref{Neff})), the rapidity distribution was calculated for the nuclear photoproduction of the $J/\psi$ in Pb-Pb collisions at $\sqrt{s} = 2.76$ TeV and $\sqrt{s} = 5.02$ TeV. Firstly, in the Figure \ref{psi1s}, our estimates are given for the centrality classes 30\%-50\%, 50\%-70\% and 70\%-90\% with $\sqrt{s} = 2.76$ TeV, using the GBW and IIM dipole models. Comparing both dipole models, there is some difference in the $|y|\gtrsim 1.0$ range although the curves shown a similar behavior. The comparison between the different centrality classes can provide more interesting information on how far the adopted formalism can be extrapoled. In especial, it was observed an increase of $\sim$ 12\% from 70\%-90\% to 50\%-70\% and of $\sim$ 13.7\% from 50\%-70\% to 30\%-50\%, for both dipole models, at $y=0$. Similarly, at $\sqrt{s}=5.02$ TeV, it was observed an increase of the $\sim$ 12\% from 70\%-90\% to 50\%-70\% and $\sim$ 13.3\% from 50\%-70\% to 30\%-50\% at $y=0$, as shown in Fig. \ref{psi1s5020}. Therefore, the relative variation between the different centrality classes is not sensitive to the increase of the energy.      

The ratio $\frac{d\sigma^{5.02}}{dy}/\frac{d\sigma^{2.76}}{dy}$ was also analysed, obtaining an increase of approximately 30\% in the central rapidity region $|y|<1.5$ for the three investigated centrality classes. This ratio is, approximately, 60\% for the same rapidity region in the UPC. It can indicate that this formalism for the effective photon flux seems less sensitive with the variation of the energy in comparison with the usual photon flux.

To compare with ALICE measurements, the average rapidity distribution was recalculated in the $2.5<y<4.0$ range, using the effective photon flux without changing the photonuclear cross section (scenario 2) and the results are presented in the Table \ref{dsigdy2}. Better agreement with the data in the central region is observed. In especial, the use of the IIM model produces better results than those produced by GBW model.   
\begin{table}[H]
\centering
\scalebox{0.7}{
\begin{tabular}{|c|c|c|c|}
\multicolumn{4}{c}{Average Rapidity Distribution - scenario 2}\tabularnewline
\hline
$d\sigma/dy$ [$\mu$b] & 30\%-50\% & 50\%-70\% & 70\%-90\%\tabularnewline
\hline 
$\text{GBW}$ & 128 & 98 & 80\tabularnewline
\hline 
$\text{IIM}$ & 107 & 80 & 67\tabularnewline
\hline 
ALICE data & $73\pm44^{+26}_{-27}\pm10$ & $58\pm16^{+8}_{-10}\pm8$ & $59\pm11^{+7}_{-10}\pm8$\tabularnewline
\hline 
\end{tabular}}
\caption{Average rapidity distribution compared with ALICE data \cite{PRL116-222301}. \label{dsigdy2}}
\end{table}

\section{The Effective Photonuclear Cross Section}

Until now, the transition from ultraperipheral to peripheral regime was performed by changing only the photon flux. However, knowing that in the built of the effective photon flux the nuclear overlap region was desconsidered, for consistency it is also necessary to discard the photon-target interaction in the overlap region. With the restriction $\Theta\left(b_1-R_A\right)$ in Eq. (\ref{glauber}) only the interaction of the photon with the non-overlap region is considered. The Eq. (\ref{glauber}) can be written as
\begin{eqnarray}
\scalebox{0.9}{$\sigma_{\textrm{dip}}^{\textrm{nucleus}}\left(x,r\right)$}&\scalebox{0.9}{$=$}&\scalebox{0.9}{$2\int dbd\alpha b\Theta(b_1-R_A)$}\nonumber
\\
&\scalebox{0.9}{$\times$}&\scalebox{0.9}{$\left\{1-\textrm{exp}\left[-\frac{1}{2}T_A(b)\sigma_{\textrm{dip}}^{\textrm{proton}}\left(x,r\right)\right]\right\}$,}\nonumber
\end{eqnarray}
where, $b_1^2=B^2+b^2+2Bb\text{cos}(\alpha)$, with $B$ the impact parameter of the nuclear collision. The combination of the modifications in the photon flux and in the photonuclear cross section constitutes the scenario 3, that produces the results for the rapidity distribution presented in the Table \ref{dsigdy3}, 
\begin{table}[H]
\centering
\scalebox{0.7}{
\begin{tabular}{|c|c|c|c|}
\multicolumn{4}{c}{Average Rapidity Distribution - scenario 3}\tabularnewline
\hline
$d\sigma/dy$ $\mu$b & 30\%-50\% & 50\%-70\% & 70\%-90\%\tabularnewline
\hline 
$\text{GBW}$ & 73 & 78 & 75\tabularnewline
\hline 
$\text{IIM}$ & 61 & 66 & 63\tabularnewline
\hline 
ALICE data & $73\pm44^{+26}_{-27}\pm10$ & $58\pm16^{+8}_{-10}\pm8$ & $59\pm11^{+7}_{-10}\pm8$\tabularnewline
\hline 
\end{tabular}}
\caption{Comparison of the results in the scenario 3 with the ALICE data \cite{PRL116-222301}. \label{dsigdy3}}
\end{table}

For completness, in addition to the $J/\psi$ state, the average rapidity distributions were also estimated for $\psi(2S)$ and for the three Y's states - $Y(1S)$, $Y(2S)$ and $Y(3S)$ with $\sqrt{s}=5.02$ TeV. All the results are summarized in the Table \ref{tudo}, where each pair of the values corresponds to GBW (left) and IIM (right) models. It can be observed that the $Y(2S)$ and $Y(3S)$ are not good discriminators, since they produce similar results for the dipole models considered in the three scenarios. It can be observed that the relative variation between the scenarios is not dependent of the dipole models (ex. $\left(S_1/S_2\right)^{GBW}\sim \left(S_1/S_2\right)^{IIM}$) for each centrality class.
\begin{table}[H]
\centering
\renewcommand{\arraystretch}{1.5}
\scalebox{0.7}{
\begin{tabular}{|c|c|c|c|}
\hline
GBW/IIM & 30\%-50\% & 50\%-70\% & 70\%-90\% \\ 
\hline
\hline
\multirow{3}{*}{$\psi(2S)$ $[\mu$b]} & \multicolumn{1}{|l|}{\textbf{S1:} 102.42/81.53} & \multicolumn{1}{|l|}{\textbf{S1:} 53.92/43.20} & \multicolumn{1}{|l|}{\textbf{S1:} 34.50/27.79} \\
& \multicolumn{1}{|l|}{\textbf{S2:} 65.51/52.31} & \multicolumn{1}{|l|}{\textbf{S2:} 51.32/41.05} & \multicolumn{1}{|l|}{\textbf{S2:} 42.45/34.02}\\
& \multicolumn{1}{|l|}{\textbf{S3:} 37.54/30.04} & \multicolumn{1}{|l|}{\textbf{S3:} 41.24/33.08} & \multicolumn{1}{|l|}{\textbf{S3:} 39.89/32.02}\\
\hline
\hline
\multirow{3}{*}{$\Upsilon(1S)$ [nb]} & \multicolumn{1}{|l|}{\textbf{S1:} 425.35/398.00} & \multicolumn{1}{|l|}{\textbf{S1:} 170.45/158.10} & \multicolumn{1}{|l|}{\textbf{S1:} 88.16/80.70} \\
& \multicolumn{1}{|l|}{\textbf{S2:} 247.7/230.86} & \multicolumn{1}{|l|}{\textbf{S2:} 184.17/171.2} & \multicolumn{1}{|l|}{\textbf{S2:} 144.45/133.87}\\
& \multicolumn{1}{|l|}{\textbf{S3:} 142.70/133.02} & \multicolumn{1}{|l|}{\textbf{S3:} 149.50/126.53} & \multicolumn{1}{|l|}{\textbf{S3:} 136.50/126.53} \\ 
\hline
\hline
\multirow{3}{*}{$\Upsilon(2S)$ [nb]} & \multicolumn{1}{|l|}{\textbf{S1:} 69.01/68.83} & \multicolumn{1}{|l|}{\textbf{S1:} 26.85/26.43} & \multicolumn{1}{|l|}{\textbf{S1:} 13.55/13.08} \\
& \multicolumn{1}{|l|}{\textbf{S2:} 39.88/39.55} & \multicolumn{1}{|l|}{\textbf{S2:} 29.51/29.17} & \multicolumn{1}{|l|}{\textbf{S2:} 23.03/22.67}\\
& \multicolumn{1}{|l|}{\textbf{S3:} 23.07/22.83} & \multicolumn{1}{|l|}{\textbf{S3:} 24.08/23.75} & \multicolumn{1}{|l|}{\textbf{S3:} 21.85/21.46} \\ 
\hline
\hline
\multirow{3}{*}{$\Upsilon(3S)$ [nb]} & \multicolumn{1}{|l|}{\textbf{S1:} 32.92/33.50} & \multicolumn{1}{|l|}{\textbf{S1:} 12.62/12.65} & \multicolumn{1}{|l|}{\textbf{S1:} 6.29/6.17} \\
& \multicolumn{1}{|l|}{\textbf{S2:} 18.95/19.17} & \multicolumn{1}{|l|}{\textbf{S2:} 14.00/14.10} & \multicolumn{1}{|l|}{\textbf{S2:} 10.90/10.93}\\
& \multicolumn{1}{|l|}{\textbf{S3:} 10.95/11.07} & \multicolumn{1}{|l|}{\textbf{S3:} 11.40/11.48} & \multicolumn{1}{|l|}{\textbf{S3:} 10.32/10.35} \\ 
\hline
\end{tabular}}
\caption{Average rapidity distribution in the region $2.5<y<4.0$ for the mesons $\psi(2S)$ and $Y(1S,2S,3S)$ for the scenarios 1,2,3 labeled by S1,S2 and S3, respectively, presented as GBW/IIM.\label{tudo}}
\end{table}

\section{$R_{AA}$ Results}\label{result}

\begin{figure}[H] 
 \centering
 \scalebox{0.5}{
 \includegraphics{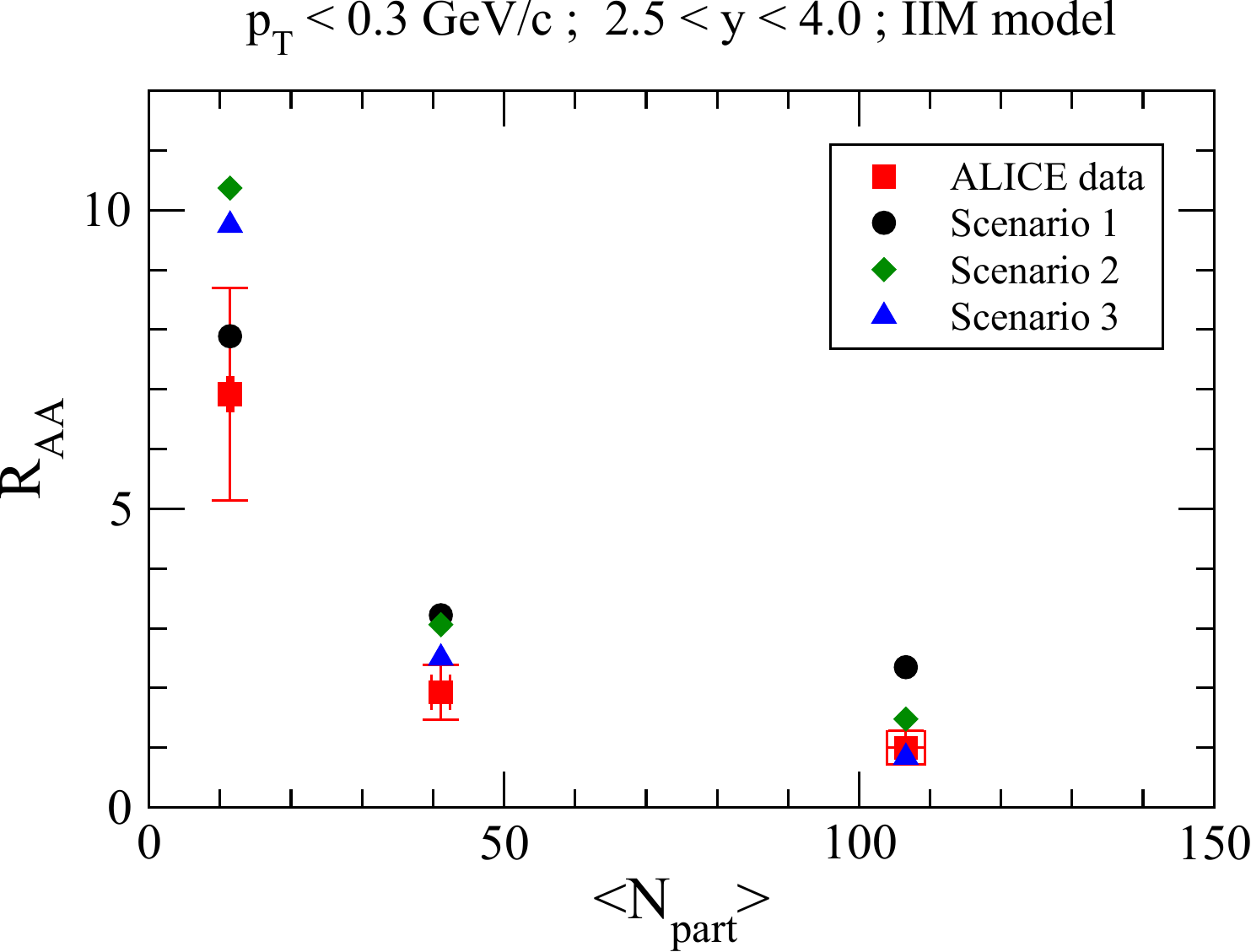}}
 \caption{Comparison of the $R_{AA}$ results with the ALICE data for the centrality classes 30\%-50\%, 50\%-70\% e 70\%-90\% \cite{PRL116-222301}.}
\label{raa}
\end{figure}

Using the Eq. (\ref{eq:raa}), the nuclear modification factor, $R_{AA}$, was calculated for the three centrality classes investigated, considering the kinematic region $p_T < 0.3$ GeV/c and $2.5<y<4.0$. Using the IIM model, which gives better results, the three scenarios developed in this paper were compared with the ALICE data, as shown in Fig. \ref{raa}. As can be observed, the scenario 1 fits with the data only in the more peripheral region where the uncertainty is higher. However, in this scenario no relevant modification was performed in relation to the ultraperipheral regime. For the scenarios 2 and 3, where a deeper dependence with $b$ was applied, better results were achieved for the more central classes where the incertainty is small. It should be considered that the ALICE measurements, which depend with the centrality of the collision, were taken following the centrality criteria developed in \cite{PRC88-044909}, where the ultraperipheral regime starts in $b\sim 20$ fm, instead of the standard $b\sim 2R_A$. Consequentely, the interval in $b$ corresponding to 70\%-90\%, for example, is not exactly the same obtained from the relation $c=b^2/4R_A$, employed in this work and closest to the Glauber model. This correction is required for a deeper comparison with the data, although main conclusions should not be affected. 

\section{Summary}

In this paper, it was calculated the average rapidity distribution for the $V=(J/\psi, \psi(2S), Y(1S), Y(2S), Y(3S))$ mesons, and the nuclear modification factor for the $J/\Psi$ state in the centrality classes 30\%-50\%, 50\%-70\% and 70\%-90\% was estimated. The ALICE data were compared with our estimates, obtained from three different approaches. In the simplest approach (scenario 1), it was obtained better aggrement with the data only in the more peripheral region, where there is a considerable uncertainty. For the more consistent approach (scenario 3), the result agrees better with the data in more central region where the incertainty is small. Although it is not yet possible to confirm that the exclusive photoproduction is fully responsible for the $J/\psi$ excess observed in ALICE, there are indications that it produces a considerable part of the effect.

\begin{acknowledgments}
We would like to thank Dr.\hspace{1mm}Ionut Arsene for usefull discussions. This work was partially financed by the Brazilian funding agency CNPq.

\end{acknowledgments}



\begin{thebibliography}{99}


\bibitem{arxiv0511047} M. Arneodo and M. Diehl, Proceedings of the Workshop on HERA and the LHC, DESY and CERN, 2004-2005.

\bibitem{EPJC54199} M. G. Ryskin, A. D. Martin and V. A. Khoze, Eur. Phys. J. C54, 199, (2008)

\bibitem{AIPCP1038-95} D.  d\textquotesingle Enterria, AIP Conf. Proc. 1038, 95, (2008).

\bibitem{PR38841} M. Diehl, Phys. Rept. 388, 41, (2003).

\bibitem{PRD94-094023} M. B. Gay Ducati, F. Kopp, M.V.T. Machado and S. Martins, Phys. Rev. D94, 094023, (2016). 

\bibitem{PRD88-017504} M. B. Gay Ducati, M. T. Griep, and M. V. T. Machado, Phys. Rev. D88, 017504, (2013).

\bibitem{JPG42-105001} G. S. dos Santos and M. V. T. Machado, J. Phys. G42, 105001, (2015). 

\bibitem{PRC87-032201} T. Lappi and H. Mantysaari, Phys. Rev. C87, 032201, (2013). 

\bibitem{PRC93-055206} V. Guzey, E. Kryshen and M. Zhalov, Phys. ReV. C93, 055206, (2016).

\bibitem{AIP1654} R. Fiore, L. Jenkovszky, V. Libov, M.V.T. Machado and A. Salii, AIP Conf. Proc. 1654, (2015).

\bibitem{TMP182-141} R. Fiore, L. Jenkovszky, V. Libov and M.V.T. Machado, Theor. Math. Phys. 182, 141-149 (2015).

\bibitem{nik} N. N. Nikolaev, B. G. Zakharov, Phys. Lett. B332, 184, (1994); Z. Phys. C64, 631, (1994).

\bibitem{NPB268-427} A. H. Mueller and J. W. Qiu, Nucl. Phys. B268, 427, (1986).

\bibitem{PLB226-167} L. N. Epele, C. A. Garcia Canal and M. B. Gay Ducati, Phys. Lett. B226, 167, (1989).

\bibitem{NPB493-305} A. L. Ayala, M. B. Gay Ducati and E. M. Levin, Nucl. Phys. B493, 305, (1997). 

\bibitem{NPB511-355} A. L. Ayala, M. B. Gay Ducati and E. M. Levin, Nucl. Phys. B511, 355, (1998). 

\bibitem{PRL116-222301} ALICE Collaboration, J. Adam et al., Phys. Rev. Lett. 116, 222301, (2016).

\bibitem{JPCS779-012039} W. Zha, J. Phys.: Conf. Series 779, 012039, (2017).

\bibitem{PRD96-056014} M. B. Gay Ducati and S. Martins, Phys. Rev. D96, 056014, (2017).

\bibitem{PRC93-044912} M. K. Gawenda and A. Szczurek, Phys. Rev. C93, 044912, (2016).

\bibitem{PRC97-044910} W. Zha \textit{et al.}, Phys. Rev. C97, 044910, (2018).

\bibitem{PLB734-314} ALICE Collaboration, B. Abelev et al., Phys. Lett. B 734, 314, (2014). 

\bibitem{PRC88-044909} ALICE Collaboration, B. Abelev et al., Phys. Rev. C88, 044909, (2013).

\bibitem{PLB718-295} ALICE Collaboration, B. Abelev et al., Phys. Lett. B 718, 295, (2012).

\bibitem{ARNPS55-271} C. A. Bertulani, S. R. Klein and J. Nystrand, Annu. Rev. Nucl. Part. Sci. 55, 271-310 (2005). 

\bibitem{PPNP39-503} F. Krauss, M. Greiner and G. Soff, Prog. Part. Nucl. Phys. 39, 503, (1997).

\bibitem{PRC14-1977} K. T. R. Davies and J. R. Nix, Phys. Rev. C 14, 1977, (1976)

\bibitem{PRC60-014903} S. Klein, J. Nystrand., Phys. Rev. C60, 014903, (1999).
             
\bibitem{PRC85-044904} A. Adeluyi and C. A. Bertulani, Phys. Rev. C 85, 044904, (2012).       

\bibitem{EPJC40-519} V. P. Gon\c calves and M. V. T. Machado, Eur. Phys. J. C 40, 519, (2005).

\bibitem{PRD74-074016} H. Kowalski, L. Motyka and G. Watt, Phys. Rev. D74, 074016, (2006).

\bibitem{PRD60-014015} A. G. Shuvaev, K. J. Golec-Biernat, A. D. Martin and M. G. Ryskin, Phys. Rev. D60, 014015, (1999).

\bibitem{PR120-1857} M. L. Good and W. D. Walker, Phys. Rev. 120, 1857, (1960).

\bibitem{ZPC75-71} J. Nemchik, N. N. Nikolaev, E. Predazzi and B. G. Zakharov, Z. Phys. C75, 71, (1997).

\bibitem{PRD90-054003} N. Armesto and A. H. Rezaeian, Phys.\ Rev.\ D90, 054003, (2014).

\bibitem{Sanda2} B.E. Cox, J.~R.~Forshaw and  R.~Sandapen, JHEP 0906, 034, (2009).

\bibitem{EPJC26-35} N. Armesto, Eur. Phys. J. C26, 35, (2002).

\bibitem{Fermi} C.W. De Jager, H. De Vries and C. De Vries, Atom. Data Nucl. Data Tabl. 14, 479, (1974).

\bibitem{PRD59-014017} K. Golec-Biernat and M. W\"usthoff, Phys. Rev. D59, 014017, (1998).

\bibitem{PLB59-199} E. Iancu, K. Itakura and S. Munier, Phys. Lett B590, 199, (2004).

\end{thebibliography}
\end{document}